# Effective field theory of Plasmas in Podolsky corrected Photonic field

Prabhat Singh and Punit Kumar

*Department of Physics, University of Lucknow, India-226007*

kumar_punit@lkouniv.ac.in

A theory for abelian plasma permeated by photons has been developed considering QED (quantum electrodynamics) generalized in Podolsky electrodynamics framework for consideration of higher order terms in electromagnetic theory. The theory traces out photonic degrees of freedom in plasma and accounts for plasma dynamics mediated by photons by calculated effective Hamiltonian. New modes of propagation have been predicted along with suppression of fields and collective behaviour. Non-Markovian behaviour is also discovered for plasma states and interactions in finite plasma system. This finds applicability in solid-state plasma, plasma confinement of magnetic and inertial nature, and laser-plasma interaction when theory is reduced to local interactions.

# 1. INTRODUCTION

More than ninety per cent of the visible universe is composed of plasma, which makes the study of plasma physics vital for the understanding of the wide range of phenomena spanning from astrophysical scale to quantum wires and quantum dots [1-4]. Apart from this plasma physics also provides insight into the universe's evolution in its early stages [5]. The need for fusion-related energy sources also makes it necessary to understand plasma behaviour very precisely [6-8]. These areas are being explored by devising particle accelerator experiments [9]. Application of theoretical understanding of non-linear phenomena in plasma has resulted in the development of laser technologies, fusion technologies and manipulation of exotic materials like quantum dots and wires [10-13]. These insights come from the models developed to study various regimes of plasmas. These models include hydrodynamic models and kinetic models and in a few cases, attempts have been made to utilize quantum field theories to develop precise models [14-16]. Fluid models though computationally simpler fail in high-energy domains or at ultra-high densities. Kinetic theories are computationally tedious and to simulate with kinetic theories as base theory comes at high computational cost [17-19]. A theory with significant accuracy, the capability of incorporating quantum effects and predicting collective behaviour with low computational cost is required.

Perturbative quantum electrodynamics (QED) gives us ease of description for plasma-electromagnetic wave interaction due to its inherent development advantage [20]. Still, it has several prerequisites which prove disadvantageous like requirements of weak coupling, slow variations, and low field strengths [21]. This requires approximations and undermines precision and accuracy. Even minor inaccuracies can result in the prediction of anomalous behaviour, such as the spurious attractive forces predicted by hydrodynamic plasma models [22]. Phenomena such as particle creation and annihilation, other quantum effects and non-linear phenomena require high precision which forces us to consider non-perturbative QED approaches [23]. This allows us to construct an effective theory where light-vacuum perturbations are expressed in terms of plasma responses to photonic fields [24-26].

For a better understanding of the collective behaviour of plasma and nonlinear effects arising in them which could be subsequently utilized in technological advancements, the present study focuses on confined plasmas (magnetically or inertially), immersed in ambient photonic fields. A theoretical framework has been developed utilizing quantum electrodynamics (QED) along with Podolsky's generalization of electrodynamics. This is marked by higher-order field derivatives. Podolsky electrodynamics inherently resolves issues arising in Maxwell's electrodynamics like that of self energies of electrons and protons and other inconsistencies. This also gives rise to new modes of propagation for electromagnetic waves in plasma and predicts novel effects [27-30].

This paper introduces a novel model for abelian plasmas [31] immersed in a uniform photonic field. The development utilizes generalized QED (QED treated with Podolsky electrodynamics) [32-35]. This results in the inclusion of higher-order electromagnetic terms to bring out the deeper dynamics of field-matter interaction. The model is applicable in various plasma environments, including solid-state plasma, magnetically and inertially confined plasma, and laser-plasma interactions. The theory derives an effective plasma Hamiltonian by eliminating photonic degrees of freedom and emphasizes plasma-specific dynamics while retaining the broader implications of field interactions. Results for the photonic field may still be obtained from the behaviour of plasma particle dynamics Section-2 introduces the Hamiltonian for charged gases and modifies it using Podolsky's generalized QED to couple it with a photonic field, resulting in a complete Hamiltonian. Section 3-traces out photonic degrees of freedom to derive an effective action and Hamiltonian for plasma systems. Section 4-applies the theory to bring out photonic observables that is brings out data encoded in effective Hamiltonian for plasma. Finally, Section-5 details the results and conclusions of the study, emphasizing its potential to advance the understanding of plasma behaviour.

## 2. SYSTEM DESCRIPTION AND BOUNDARY CONDITIONS

We consider a completely ionized gas subsystem composed of positive charges (massive) and electrons. Since, electrons are lighter than ions (positive charges), their motion would be significantly faster. For all practical purposes, we may consider ions to be static in comparison to electrons [36]. All the processes for electrons would be on a much smaller time scale than that for ions. Thus, ions could be considered at rest while studying the dynamics of electrons. This may also be considered in the form of negative charges surrounding a non-dynamical core of a positive charge.

A gas of charged particles with a neutralizing background containing a non-dynamical core may be expressed in terms of Hamiltonian of the following form,

$$H_{ionised} = \sum_j \left( \frac{\hat{p}_{e_j}^2}{2m_e} \right) + \hat{H}_{ee} + \hat{H}_{ep} + \hat{H}_{pp} + \sum_j \left( \frac{\hat{p}_{p_j}^2}{2m_p} \right), \tag{1}$$

where, $\hat{p}_{e_j}^2$ and $\hat{p}_{p_j}^2$ are canonical momentum of the $j^{th}$ electron and proton, $H_{ionised}$ is total Hamiltonian for charged gas with $\sum_j \left( \frac{\hat{p}_{e_j}^2}{2m_e} \right)$ and $\sum_j \left( \frac{\hat{p}_{p_j}^2}{2m_p} \right)$ being description of free electron and proton respectively. $H_{ee}$ is Hamiltonian for interaction of electrons and $\hat{H}_{ep}, \hat{H}_{pp}$ are the Hamiltonian for interaction of electrons and protons and proton and protons respectively.

## 2.1. Podolsky modifications in electromagnetic parameters and Hamiltonian.

Generalized electrodynamics also termed as Podolsky's electrodynamics introduces higher order terms in field equations. This is done by involving higher order corrections in the lagrangian of electromagnetic field. Introduction of a parameter '$a$' with dimensions of length and in some cases '$m$' with dimensions of mass called 'Podolsky length' and 'Podolsky mass' respectively enables this generalization. The Lagrangian density in Podolsky electrodynamics is given by [37,38],

$$L = -\frac{1}{4} F_{\mu\nu} F^{\mu\nu} + \frac{a^2}{2} \partial_\lambda F^{\lambda\mu} \partial^\nu F_{\nu\mu},$$  (2)

where, $F_{\mu\nu} = \partial_\mu A_\nu - \partial_\nu A_\mu$ is the field strength tensor. $A_\mu \left(= \mathbf{A}, A_0\right)$ is the four-potential ($\mathbf{A}$ is the vector potential and $A_0$ is the scalar potential), and $a$ is 'Podolsky length', which characterizes the scale of the higher-order corrections mentioned earlier. The first term represents standard Maxwellian Lagrangian while the second term represents Podolsky corrections.

### 2.1.1. Vector potential

The field equations derived from the Lagrangian in equation (2) is [38],

$$\left(1 - a^2 \Box\right) \partial_\mu F^{\mu\nu} = \mu_0 J^\nu,$$  (3)

where, $\Box = \partial_\mu \partial^\mu$ is the d'Alembertian operator and $J^\nu (= \rho, \mathbf{J}))$ is the four-current. $\rho$ is the charge density and $\mathbf{J}$ is current density. $\mu, \nu$ and $\lambda$ are standard indices of electrodynamics [39]. For the spatial components, equation (3) simplifies to,

$$(1-a^2)\left(\Box \mathbf{A} - \nabla(\nabla \cdot \mathbf{A}) - \frac{\partial \nabla A_0}{\partial t}\right) = \mu_0 \mathbf{J}. \tag{4}$$

We choose to work in the Lorenz gauge for the equation to be relativistically invariant. Lorenz gauge in Podolsky electrodynamics [40] is stated as,

$$\partial_\mu A^\mu + a^2 \Box \partial_\mu A^\mu = 0. \tag{5}$$

In this gauge, the equation for vector potential becomes,

$$(1-a^2)(\Box \mathbf{A}) = \mu_0 \mathbf{J}. \tag{6}$$

We see that the solution of this equation will give us modified vector potential that will be further utilized in calculating magnetic field and Hamiltonian for electromagnetic wave. To solve this we utilize Fourier transform, which for the vector potential $\mathbf{A}(\mathbf{r}, t)$ is [41],

$$\mathbf{A}(\mathbf{r},t) = \int \frac{d^3k}{(2\pi)^3} \int \frac{d\omega}{(2\pi)^3} \tilde{\mathbf{A}}(\mathbf{k}, \omega) e^{i(\mathbf{k} \cdot \mathbf{r} - \omega t)}. \tag{7}$$

Fourier transform allows expression of vector potential in the form of summation of different fundamental modes of electromagnetic waves, each mode being a plane wave. It also expresses different modes arising due to standard contributions and Podolsky contribution separately. To solve the integral, we use the residue theorem considering the contribution from the poles in the $\omega$ plane. Each pole represent different mode of propagation. The poles are located at,

$$\omega = \pm |\mathbf{k}| \text{ and } \omega = \pm \frac{|\mathbf{k}|}{\sqrt{1+a^2 k^2}}. \tag{8}$$

These correspond to different modes of propagation,

Standard Mode: $\omega = \pm |\mathbf{k}|$,

and

Podolsky Mode: $\omega = \pm \frac{|\mathbf{k}|}{\sqrt{1+a^2 k^2}}$.

We can see that for the standard mode, the contribution is similar to the solution in classical electrodynamics [42,43], its expansion in terms of Fourier modes for our specific case will take the form,

$$\mathbf{A}(\mathbf{r},t) = \int \frac{d^3 k}{(2\pi)^3} \sum_\lambda \left[ \mathbf{e}_{\mathbf{k},\lambda} A_{\mathbf{k},\lambda}(t) e^{i\mathbf{k}\cdot\mathbf{r}} + c.c \right], \tag{9}$$

where, $\mathbf{e}_{\mathbf{k},\lambda}$ are the polarization vectors, $A_{\mathbf{k},\lambda}$ are the time-dependent Fourier coefficients, $\lambda$ labels the polarization states and c.c denotes complex conjugate of the previous term. For Podolsky electrodynamics, the Fourier coefficients $A_{\mathbf{k},\lambda}$ satisfy a modified dispersion relation in equation (9). The modified dispersion relation, incorporating the Podolsky length scale $a$, is,

$$(1+a^2 \mathbf{k}^2)\omega_\mathbf{k}^2 = c^2 \mathbf{k}^2. \tag{10}$$

In the Lorenz gauge, the standard and Podolsky vector potentials are expressed as,

$$\mathbf{A}_{standard}(\mathbf{r},t) = \int \frac{d^3k}{(2\pi)^3} \sum_\lambda \left( \mathbf{e}_{\mathbf{k},\lambda} A_{\mathbf{k},\lambda} e^{i(\mathbf{k}\cdot\mathbf{r}-\omega_k t)} + c.c \right),$$
(11)

and

$$\mathbf{A}_{Podolsky}(\mathbf{r},t) = \int \frac{d^3k}{(2\pi)^3} \sum_\lambda \left( \mathbf{e}_{\mathbf{k},\lambda} A^{Podolsky}_{\mathbf{k},\lambda} e^{i(\mathbf{k}\cdot\mathbf{r}-\omega^{Podolsky}_{\mathbf{k}} t)} + c.c. \right).$$
(12)

The standard dispersion relation is $\omega = \pm c|\mathbf{k}|$ and Podolsky dispersion relation is $\omega^{Podolsky} = \pm \frac{c|\mathbf{k}|}{\sqrt{1+a^2 k^2}}$ where, $A_{\mathbf{k},\lambda}$ and $A^{Podolsky}_{\mathbf{k},\lambda}$ are the Fourier components of the vector potentials. The Fourier coefficients $A^{standard}_{\mathbf{k},\lambda}$ and $A^{Podolsky}_{\mathbf{k},\lambda}$ are related by the fact that in Podolsky electrodynamics, the modification typically introduces a factor depending on the characteristic length scale $a$,

$$A^{Podolsky}_{\mathbf{k},\lambda} = \frac{A^{standard}_{\mathbf{k},\lambda}}{\sqrt{1+a^2 \mathbf{k}^2}}.$$
(13)

The total vector potential results from combination of both the contributions is,

$$\mathbf{A}_{total}(\mathbf{r},t) = \int \frac{d^3k}{(2\pi)^3} \sum_\lambda \left( \mathbf{e}_{\mathbf{k},\lambda} \left[ A^{standard}_{\mathbf{k},\lambda} + A^{Podolsky}_{\mathbf{k},\lambda} \right] e^{i(\mathbf{k}\cdot\mathbf{r}-\omega_{\mathbf{k},total} t)} + c.c. \right),$$
(14)

where, $\omega_{\mathbf{k},total}$ is the effective frequency incorporating both contributions. For simplicity, we express the total vector potential as,

$$\mathbf{A}_{total}(\mathbf{r},t) = \int \frac{d^3k}{(2\pi)^3} \sum_\lambda \mathbf{e}_{\mathbf{k},\lambda} \left[ \frac{A^{standard}_{\mathbf{k},\lambda}}{\sqrt{1+a^2 \mathbf{k}^2}} \right] e^{i(\mathbf{k}\cdot\mathbf{r}-\omega_{\mathbf{k},total} t)} + c.c..$$
(15)

To quantize the vector potential, we replace the Fourier coefficients with creation and annihilation operators [44],

$$\hat{\mathbf{A}}_{total}(\mathbf{r},t) = \int \frac{d^3k}{(2\pi)^3} \sum_{\lambda} \left( \mathbf{e}_{\mathbf{k},\lambda} \left[ \frac{\hat{a}_{\mathbf{k},\lambda}}{\sqrt{1+a^2\mathbf{k}^2}} \right] e^{i(\mathbf{k}\cdot\mathbf{r}-\omega_{\mathbf{k},total}t)} + \mathbf{e}^*_{\mathbf{k},\lambda} \left[ \frac{\hat{a}^\dagger_{\mathbf{k},\lambda}}{\sqrt{1+a^2\mathbf{k}^2}} \right] e^{-i(\mathbf{k}\cdot\mathbf{r}-\omega_{\mathbf{k},total}t)} \right), \quad (16)$$

where, $\hat{a}_{\mathbf{k},\lambda}$ and $\hat{a}^\dagger_{\mathbf{k},\lambda}$ are the annihilation and creation operators satisfying the commutation relations [45],

$$\left[ \hat{a}_{\mathbf{k},\lambda}, \hat{a}^\dagger_{\mathbf{k},\lambda} \right] = (2\pi)^3 \delta^3(\mathbf{k}-\mathbf{k}')\delta_{\lambda\lambda'}. \quad (17)$$

This ensures that creation and annihilation is meaningful only when they are of same modes. The final quantized expression for the total vector potential in Podolsky electrodynamics becomes,

$$\hat{\mathbf{A}}_{total}(\mathbf{r},t) = \int \frac{d^3k}{(2\pi)^3} \sum_{\lambda} \left( \mathbf{e}_{\mathbf{k},\lambda} \left[ \frac{\hat{a}_{\mathbf{k},\lambda}}{\sqrt{1+a^2\mathbf{k}^2}} \right] e^{i\left(\mathbf{k}\cdot\mathbf{r}-\frac{c|\mathbf{k}|}{\sqrt{1+a^2\mathbf{k}^2}}t\right)} + h.c. \right), \quad (18)$$

where, $\mathbf{e}_{\mathbf{k},\lambda}$ are the polarization vectors, $\frac{1}{\sqrt{1+a^2\mathbf{k}^2}}$ is the Podolsky correction factor, $\omega_{\mathbf{k},total} = \frac{c|\mathbf{k}|}{\sqrt{1+a^2\mathbf{k}^2}}$ is the modified dispersion relation incorporating Podolsky's theory. It is not the direct sum of the standard and Podolsky mode but representation of one dominant mode that would effectively represent combined effect and h.c stands for hermitian conjugate.

We further, isolate mode function for vector potential. The mode functions here are the components of the plane wave solutions $e^{i(\mathbf{k}\cdot\mathbf{r}-\omega_{\mathbf{k},total}t)}$ and $e^{-i(\mathbf{k}\cdot\mathbf{r}-\omega_{\mathbf{k},total}t)}$, which describe how the electromagnetic field propagates in space and time. The mode functions in the Hamiltonian describe the quantum states of the electromagnetic field that contain regularizing effects of Podolsky electrodynamics. This puts constraints on the field due to which it behaves properly even at high energies that is does not result in divergent solutions for self energies of point charges. These functions describe how photons are created, propagate, and interact with charged particles present in the system. The mode functions for the vector potential are the terms that are present alongside the creation and annihilation operators,

$$f_{\mathbf{k},\lambda}(\mathbf{r},t) = \frac{1}{\sqrt{1+a^2\mathbf{k}^2}} e^{i(\mathbf{k}\cdot\mathbf{r}-\omega_{\mathbf{k},total}t)}. \tag{19}$$

For the conjugate mode function associated with the creation operator,

$$f^*_{\mathbf{k},\lambda}(\mathbf{r},t) = \frac{1}{\sqrt{1+a^2\mathbf{k}^2}} e^{-i(\mathbf{k}\cdot\mathbf{r}-\omega_{\mathbf{k},total}t)}. \tag{20}$$

### 2.1.2. Magnetic field

Classical electrodynamics establishes that the magnetic field $\mathbf{B}(\mathbf{r},t)$ is related to the vector potential $\mathbf{A}(\mathbf{r},t)$ by the curl. To obtain the quantized magnetic field, we apply the curl operator to the quantized vector potential,

$$\hat{\mathbf{B}}_{total}(\mathbf{r},t) = \nabla \times \hat{\mathbf{A}}_{total}(\mathbf{r},t). \tag{21}$$

Expressing the curl in Fourier Space, we can utilize the identity [46],

$$\nabla \times \left(\mathbf{e}_{\mathbf{k},\lambda} e^{i\mathbf{k}\cdot\mathbf{r}}\right) = i\mathbf{k} \times \mathbf{e}_{\mathbf{k},\lambda} e^{i\mathbf{k}\cdot\mathbf{r}}. \tag{22}$$

Therefore,

$$\hat{\mathbf{B}}_{total}(\mathbf{r},t) = \nabla \times \hat{\mathbf{A}}_{total}(\mathbf{r},t) = \int \frac{d^3k}{(2\pi)^3} \sum_\lambda (i\mathbf{k} \times \mathbf{e}_{\mathbf{k},\lambda}) \left( \left[ \frac{\hat{a}_{\mathbf{k},\lambda}}{\sqrt{1+a^2\mathbf{k}^2}} \right] e^{i(\mathbf{k}\cdot\mathbf{r}-\omega_{\mathbf{k},total}t)} - \left[ \frac{\hat{a}^\dagger_{\mathbf{k},\lambda}}{\sqrt{1+a^2\mathbf{k}^2}} \right] e^{-i(\mathbf{k}\cdot\mathbf{r}-\omega_{\mathbf{k},total}t)} \right). \quad (23)$$

The sign change in the second term arises from the derivative acting on the complex conjugate exponential. Combining the above steps, the quantized total magnetic field $\hat{\mathbf{B}}_{total}(\mathbf{r},t)$ becomes,

$$\hat{\mathbf{B}}_{total}(\mathbf{r},t) = \int \frac{d^3k}{(2\pi)^3} \sum_\lambda \left[ i\mathbf{k} \times \mathbf{e}_{\mathbf{k},\lambda} \frac{\hat{a}_{\mathbf{k},\lambda}}{\sqrt{1+a^2\mathbf{k}^2}} e^{i(\mathbf{k}\cdot\mathbf{r}-\omega_{\mathbf{k},total}t)} - i\mathbf{k} \times \mathbf{e}^*_{\mathbf{k},\lambda} \frac{\hat{a}^\dagger_{\mathbf{k},\lambda}}{\sqrt{1+a^2\mathbf{k}^2}} e^{-i(\mathbf{k}\cdot\mathbf{r}-\omega_{\mathbf{k},total}t)} \right]. \quad (24)$$

To make the expression further, more compact, we define

$$\mathbf{f}_{\mathbf{k},\lambda} = i\mathbf{k} \times \mathbf{e}_{\mathbf{k},\lambda}, \quad (25)$$

and

$$\mathbf{f}^*_{\mathbf{k},\lambda} = -i\mathbf{k} \times \mathbf{e}^*_{\mathbf{k},\lambda}. \quad (26)$$

Thus, the expression (24) becomes,

$$\hat{\mathbf{B}}_{total}(\mathbf{r},t) = \int \frac{d^3k}{(2\pi)^3} \sum_\lambda \left[ \mathbf{f}_{\mathbf{k},\lambda} \frac{\hat{a}_{\mathbf{k},\lambda}}{\sqrt{1+a^2\mathbf{k}^2}} e^{i(\mathbf{k}\cdot\mathbf{r}-\omega_{\mathbf{k},total}t)} + \mathbf{f}^*_{\mathbf{k},\lambda} \frac{\hat{a}^\dagger_{\mathbf{k},\lambda}}{\sqrt{1+a^2\mathbf{k}^2}} e^{-i(\mathbf{k}\cdot\mathbf{r}-\omega_{\mathbf{k},total}t)} \right], \quad (27)$$

where, $\mathbf{e}_{\mathbf{k},\lambda}$, the polarization vectors are orthogonal to $\mathbf{k}$ that is $\mathbf{e}_{\mathbf{k},\lambda}\cdot\mathbf{k} = 0$ and are also orthogonal to each other i.e. $\mathbf{e}_{\mathbf{k},\lambda}\cdot\mathbf{e}_{\mathbf{k},\lambda'} = \delta_{\lambda\lambda'}$. The mode functions for the magnetic field are,

$$h_{\mathbf{k},\lambda}(\mathbf{r},t) = \frac{i\mathbf{k} \times \mathbf{e}_{\mathbf{k},\lambda}}{\sqrt{1+a^2\mathbf{k}^2}} e^{i(\mathbf{k}\cdot\mathbf{r}-\omega_{\mathbf{k},total}t)}, \quad (28)$$

and for the conjugate part,

$$h^*_{\mathbf{k},\lambda}(\mathbf{r},t) = \frac{-i\mathbf{k}\times\mathbf{e}_{\mathbf{k},\lambda}}{\sqrt{1+a^2\mathbf{k}^2}} e^{-i(\mathbf{k}\cdot\mathbf{r}-\omega_{\mathbf{k},total}t)}. \tag{29}$$

These mode functions describe how the magnetic field propagates and interacts with the quantized photon modes.

Annihilation and creation operators $\hat{a}_{\mathbf{k},\lambda}$ and $\hat{a}^\dagger_{\mathbf{k},\lambda}$ destroys and creates a photon in the mode characterized by $\mathbf{k}$ and $\lambda$ respectively. These also satisfy the commutation relation [47],

$$\left[\hat{a}_{\mathbf{k},\lambda}, \hat{a}_{\mathbf{k}',\lambda'}\right] = \left[\hat{a}^\dagger_{\mathbf{k},\lambda}, \hat{a}^\dagger_{\mathbf{k}',\lambda'}\right] = 0. \tag{30}$$

Podolsky modification factor results in a momentum-dependent attenuation, this regularizes the field at high momenta (large $\mathbf{k}$) which reflects the non-local behaviour or finite range of interactions due the characteristic length scale $a$.

### 2.1.3. Hamiltonian for electromagnetic field

Podolsky's theory introduces higher-derivative terms into the Lagrangian of electromagnetic field, which leads to a modified Hamiltonian density,

$$H_{EM-Podolsky} = \frac{1}{2}\left(\mathbf{E}^2 - \mathbf{B}^2\right) + \frac{a^2}{2}\left[(\nabla\cdot\mathbf{E})^2 + (\nabla\times\mathbf{B})^2\right], \tag{31}$$

where, $a$ is the characteristic length scale of Podolsky's theory, $\mathbf{E}$ and $\mathbf{B}$ are electric and magnetic field respectively. The total Hamiltonian is the integral of the Hamiltonian density over all space,

$$H_{EM-Podolsky-total} = \int d^3\mathbf{r}\, H_{EM-Podolsky}. \tag{32}$$

In the Lorenz gauge when there is no spatial variation and $\hat{\mathbf{E}}$ depends only on time variation of vector potential, the electric field $\hat{\mathbf{E}}(\mathbf{r},t)$ is related to the vector potential by,

$$\hat{\mathbf{E}}(\mathbf{r},t) = -\frac{\partial \hat{\mathbf{A}}(\mathbf{r},t)}{\partial t}. \tag{33}$$

Substituting the quantized expression for $\hat{\mathbf{A}}_{total}$,

$$\hat{\mathbf{E}}_{total}(\mathbf{r},t) = i\int \frac{d^3k}{(2\pi)^3} \sum_\lambda \mathbf{e}_{\mathbf{k},\lambda} \frac{\hat{a}_{\mathbf{k},\lambda} \omega_{\mathbf{k},total}}{\sqrt{1+a^2\mathbf{k}^2}} e^{i(\mathbf{k}\cdot\mathbf{r}-\omega_{\mathbf{k},total}t)} - h.c., \tag{34}$$

The mode functions for the electric field here are,

$$g_{\mathbf{k},\lambda}(\mathbf{r},t) = \frac{-i\omega_{\mathbf{k},total}}{\sqrt{1+a^2\mathbf{k}^2}} e^{i(\mathbf{k}\cdot\mathbf{r}-\omega_{\mathbf{k},total}t)}, \tag{35}$$

and for the conjugate part,

$$g^*_{\mathbf{k},\lambda}(\mathbf{r},t) = \frac{i\omega_{\mathbf{k},total}}{\sqrt{1+a^2\mathbf{k}^2}} e^{-i(\mathbf{k}\cdot\mathbf{r}-\omega_{\mathbf{k},total}t)}. \tag{36}$$

Substituting the values of vector potential, magnetic field and calculated electric field from equations (18), (27) and (34) into equation (31) we get quantized hamiltonian for electromagnetic wave in Podolsky theory,

$$H_{EM} = \frac{d^3k}{(2\pi)^3} \sum_\lambda \hbar \omega_{k,total} \left( \hat{a}^\dagger_{k,\lambda} \hat{a}_{k,\lambda} + \frac{1}{2} \right), \tag{37}$$

where, $\hbar\omega_{\mathbf{k},total}$ is the energy of a photon in the mode characterized by $\mathbf{k}$ and $\lambda$, with the modified dispersion relation $\omega_{\mathbf{k},total} = \dfrac{c|\mathbf{k}|}{\sqrt{1+a^2\mathbf{k}^2}}$. Total Hamiltonian for electromagnetic wave comes out to be,

$$H_{EM} = \int \frac{d^3k}{(2\pi)^3} \sum_\lambda \hbar \frac{c|k|}{\sqrt{1+a^2k^2}} \left( \hat{a}^\dagger_{k,\lambda} \hat{a}_{k,\lambda} + \frac{1}{2} \right). \qquad (38)$$

The Hamiltonian represents the sum of the energy of all possible photon modes, each mode contributing according to the modified dispersion relation. The factor $\dfrac{1}{\sqrt{1+a^2\mathbf{k}^2}}$ reduces the contribution of high-momentum modes, leading to regularization of the field energy. Finite length scale of the characteristic length scale $a$ introduces a cutoff, preventing the ultraviolet divergences typical in classical electrodynamics. The term $\dfrac{1}{2}\hbar\omega_{\mathbf{k},total}$ represents the zero-point energy of each mode, even in the vacuum state.

## 2.2. Hamiltonian of plasma in cavity

The total Hamiltonian for the system, including the free electron and proton Hamiltonians, the electromagnetic field Hamiltonian, and the interaction Hamiltonians is,

$$\hat{H}_{total} = \hat{H}_{free-electron} + \hat{H}_{free-proton} + \hat{H}_{e-e} + \hat{H}_{e-p} + \hat{H}_{p-p} + \hat{H}_{EM}, \qquad (39)$$

where, $\hat{H}_{free-electron}$ is the free electron Hamiltonian, including minimal coupling to the electromagnetic field. $\hat{H}_{free-proton}$ is the free proton Hamiltonian. $\hat{H}_{e-e}, \hat{H}_{e-p}$, and $\hat{H}_{p-p}$ are the interaction Hamiltonians as described above. $\hat{H}_{EM}$ is the electromagnetic field Hamiltonian (modified by Podolsky electrodynamics). The interactions between plasma particles mediated by photons have also been modified by Podolsky electromagnetic theory.

### 2.2.1. Free particle Hamiltonian

The free electron Hamiltonian is,

$$\hat{H}_{free-electron} = \sum_{j=1}^{N_e} \frac{1}{2m_e}\left[\hat{\mathbf{p}}_j^2 - 2e\hat{\mathbf{p}}_j \cdot \hat{\mathbf{A}}_{total}^2(\hat{\mathbf{r}}_j,t) + e^2\hat{\mathbf{A}}_{total}^2(\hat{\mathbf{r}}_j,t)\right] + eA_0(\hat{\mathbf{r}}_j,t) \tag{40}$$

where, $\hat{\mathbf{p}}_j = -i\hbar\nabla$ is the momentum operator for the $j^{th}$ electron. Free Proton Hamiltonian (with minimal coupling),

$$\hat{H}_{free-proton} = \sum_{i=1}^{N_p} \frac{1}{2m_p}\left[\hat{\mathbf{p}}_i^2 - 2e\hat{\mathbf{p}}_i \cdot \hat{\mathbf{A}}_{total}^2(\hat{\mathbf{r}}_i,t) + e^2\hat{\mathbf{A}}_{total}^2(\hat{\mathbf{r}}_i,t)\right] + eA_0(\hat{\mathbf{r}}_i,t) \tag{41}$$

where, $\hat{\mathbf{A}}_{total}$ is the total vector potential in Podolsky electrodynamics. This term accounts for the interaction of free electrons with photons in the quantized electromagnetic field. $eA_0(\mathbf{r}_j,t)$ represents the interaction with the scalar potential, where $A_0(\mathbf{r},t)$ can be determined based on the gauge choice but since there is no spatial variation in field that is field is uniform this term can be neglected. In the Lorenz gauge, this scalar potential can describe Coulomb-like effects if needed.

### 2.2.2. Interaction terms

We now introduce the interaction Hamiltonians, modified to incorporate Podolsky regularization and Debye shielding [48].

*2.2.2.1. Electron-Electron Interaction.* Starting from Podolsky modified Poisson equation, substituting expression for charge density from Boltzmann distribution, defining Debye length and obtaining solution for Poisson's equation gives us potential. The potential between two electrons at positions $\hat{\mathbf{r}}_j$ and $\hat{\mathbf{r}}_k$ is,

$$V_{e-e}(r) = \frac{e^2}{r}\left(1 - \frac{r^2}{a^2}\right)e^{\frac{-r}{\lambda_D}}, \tag{42}$$

where, $r = |\hat{\mathbf{r}}_j - \hat{\mathbf{r}}_k|$. The corresponding Hamiltonian is,

$$\hat{H}_{e-e} = \frac{1}{2}\sum_{j\neq k}^{N_e} e_j e_k V_{e-e}\left(|\hat{\mathbf{r}}_j - \hat{\mathbf{r}}_k|\right). \tag{43}$$

*2.2.2.2. Electron-Proton Interaction.* The interaction between electrons and protons is attractive, for which the potential is given by,

$$V_{p-e}(r) = \frac{e^2}{r}\left(1 - \frac{r^2}{a^2}\right)e^{\frac{-r}{\lambda_D}}. \tag{44}$$

The corresponding Hamiltonian is,

$$\hat{H}_{e-p} = \frac{1}{2}\sum_{j=1}^{N_e}\sum_{i=1}^{N_p} \left(-e_j e_i V_{e-p}\left(|\hat{\mathbf{r}}_j - \hat{\mathbf{r}}_i|\right)\right). \tag{45}$$

*2.2.2.3. Proton-Proton Interaction.* Similar to electron-electron interaction, but for protons,

$$\hat{H}_{p-p} = \frac{1}{2}\sum_{i\neq j}^{N_p} e_i e_j V_{p-p}\left(|\hat{\mathbf{r}}_i - \hat{\mathbf{r}}_j|\right), \tag{46}$$

where, $V_{p-p}(\mathbf{r})$ has the same form as $V_{e-e}(\mathbf{r})$, with $e_p$ replacing $e_e$.

### 2.2.3. Total Hamiltonian

Summing all these contributions, the total Hamiltonian for the system is,

$$\hat{H}_{total} = \hat{H}_{free-electron} + \hat{H}_{free-proton} + \hat{H}_{e-e} + \hat{H}_{e-p} + \hat{H}_{p-p} + \hat{H}_{EM}.$$

Substituting values from equation (40), (41), (43), (45), (46) and (38) in equation (39) we get,

$$\hat{H}_{total} = \sum_{j=1}^{N_e} \frac{1}{2m_e} \left[ \hat{p}_j^2 - 2e\hat{p}_j \cdot \hat{\mathbf{A}}_{total}(\hat{\mathbf{r}}_j, t) + e^2 \hat{\mathbf{A}}_{total}^2(\hat{\mathbf{r}}_j, t) \right]$$

$$+ \sum_{i=1}^{N_p} \frac{1}{2m_p} \left[ \hat{p}_i^2 + -2e\hat{p}_i \cdot \hat{\mathbf{A}}_{total}(\hat{\mathbf{r}}_i, t) + e^2 \hat{\mathbf{A}}_{total}^2(\hat{\mathbf{r}}_i, t) \right]$$

$$+ \frac{1}{2} \sum_{j \neq k}^{N_e} \frac{e^2}{|\hat{\mathbf{r}}_j - \hat{\mathbf{r}}_k|} \left( 1 - \frac{|\hat{\mathbf{r}}_j - \hat{\mathbf{r}}_k|^2}{a^2} \right) e^{\frac{-|\hat{\mathbf{r}}_j - \hat{\mathbf{r}}_k|}{\lambda_D}}$$

$$+ \sum_{j=1}^{N_e} \sum_{i=1}^{N_p} \frac{e^2}{|\hat{\mathbf{r}}_j - \hat{\mathbf{r}}_i|} \left( 1 - \frac{|\hat{\mathbf{r}}_j - \hat{\mathbf{r}}_i|^2}{a^2} \right) e^{\frac{-|\hat{\mathbf{r}}_j - \hat{\mathbf{r}}_i|}{\lambda_D}}$$

$$+ \frac{1}{2} \sum_{i \neq j}^{N_p} \frac{e^2}{|\hat{\mathbf{r}}_i - \hat{\mathbf{r}}_j|} \left( 1 - \frac{|\hat{\mathbf{r}}_i - \hat{\mathbf{r}}_j|^2}{a^2} \right) e^{\frac{-|\hat{\mathbf{r}}_i - \hat{\mathbf{r}}_j|}{\lambda_D}}$$

$$+ \int \frac{d^3k}{2\pi} \sum_\lambda \hbar \frac{c|\mathbf{k}|}{\sqrt{1 + a^2 \mathbf{k}^2}} \left( \hat{a}_{k,\lambda}^\dagger \hat{a}_{k,\lambda} + \frac{1}{2} \right). \tag{48}$$

This is the full Hamiltonian for the plasma system with electrons and protons interacting in the Podolsky electrodynamics framework with Debye screening.

### 2.2.4. Hamiltonian for terms quadratically dependent on Bosonic operators

The terms containing annihilation and creation operators in pair that is they have quadratic dependence on bosonic operators give description of photons, their energy, motion and enunciate how interaction of photons and plasma particles take place. This in turn describes scattering, emission, and absorption processes, the renormalization and stabilization of the field due to Podolsky regularization. These phenomena circumvent the typical UV divergences of standard QED. Purpose of isolating these quadratic terms, is to gain insight into the role of the quantized photon field and its interaction with matter. This is essential for understanding radiative corrections, photon propagation, and the fundamental quantum behaviour of the electromagnetic field in Podolsky electrodynamics.

We now derive the mode-mode coupling constant $\Delta_{\kappa,\kappa'}$ and the mode interaction function $\hat{U}_{\kappa,\kappa'}$. These represent the photon-photon interactions mediated by the plasma and the mode coupling induced by the presence of charged particles.

*2.2.4.1. Derivation of $\Delta_{\kappa,\kappa'}$ (Mode-Mode Coupling Constant).* The term $\Delta_{\kappa,\kappa'}$ describes the coupling strength between photon modes $\kappa$ and $\kappa'$ due to their interaction with the plasma particles (electrons and protons). To calculate $\Delta_{\kappa,\kappa'}$, we account for collective plasma effects, such as plasma oscillations and screening, photon-plasma interactions, which can lead to pair creation/annihilation or mode mixing in the plasma.

In the plasma, the interaction between charged particles is screened by the presence of free charges, leading to a screened Coulomb potential. The Fourier transform of the screened potential $V$ gives the momentum-space representation of the interaction,

$$V_{screened}(\mathbf{q}) = \frac{4\pi e^2}{\left(|\mathbf{q}|^2 + \frac{1}{\lambda_D^2}\right)} \left(1 - \frac{1}{a^2} \frac{2\frac{1}{\lambda_D} - |\mathbf{q}|^2}{|\mathbf{q}|^2 + \frac{1}{\lambda_D^2}}\right),$$
(49)

where, $\mathbf{q} = \mathbf{k} - \mathbf{k}'$ is the momentum transfer between two photon modes and $\lambda_D$ is the Debye length.

The mode-mode coupling constant $\Delta_{\kappa,\kappa'}$ is the Fourier transform of the screened potential $V$ described before. The coupling constant $\Delta_{\kappa,\kappa'}$ also accounts for the Podolsky regularization, which modifies the momentum-space structure of the interaction. Thus, the exact expression for $\Delta_{\kappa,\kappa'}$ is,

$$\Delta_{\kappa,\kappa'} = \frac{4\pi e^2}{\left(|\mathbf{q}|^2 + \frac{1}{\lambda_D^2}\right)} \left(1 - \frac{1}{a^2} \frac{2\frac{1}{\lambda_D} - |\mathbf{q}|^2}{|\mathbf{q}|^2 + \frac{1}{\lambda_D^2}}\right) \frac{1}{\sqrt{(1 + a^2|\mathbf{k}|^2)(1 + a^2|\mathbf{k}'|^2)}}.$$
(50)

Podolsky regularization smoothens the high-momentum contributions and prevent divergences at short distances. This structure of mode coupling constant predicts modified plasma oscillations due to modified screening. This also predicts enhanced stability especially in case of high energy and high density plasma. Dependence on $|\mathbf{q}|^2$ also suggests anisotropic behaviour in mode coupling. This directional dependence may lead to topological effects which have not previously been predicted. Podolsky parameter also suppresses small scale fluctuations. Dependence on wave vectors in denominator may facilitate mode locking, energy cascades and wave synchronization. This also points towards screened plasma clusters in certain conditions.

*2.2.4.2. Derivation of $\hat{U}_{\kappa,\kappa'}$ (Mode Interaction Function.* The term $\hat{U}_{\kappa,\kappa'}$ captures the spatial dependence of the interaction between photon modes $\kappa$ and $\kappa'$, which is mediated by the charged particles in the plasma. To derive $\hat{U}_{\kappa,\kappa'}$, we consider how the electron and proton positions affect the interaction between the photon modes. $\hat{U}_{\kappa,\kappa'}$ is determined by the spatial overlap between the photon modes and the density fluctuations in the plasma. The electron and proton positions introduce spatial variations in the photon field, which lead to mode mixing. The general form of $\hat{U}_{\kappa,\kappa'}$ depends on the electron density fluctuations and the wave functions of the photon modes.

The charge density $\rho(\mathbf{r})$ in plasma is given by the sum of the electron and proton charge densities. For $N_e$ electrons and $N_p$ protons, the quantized charge density is,

$$\rho(\mathbf{r}) = \sum_{j=1}^{N_e} e\delta(\mathbf{r}-\hat{\mathbf{r}}_j) - \sum_{i=1}^{N_p} e\delta(\mathbf{r}-\hat{\mathbf{r}}_i), \tag{51}$$

where, $\hat{\mathbf{r}}_j$ and $\hat{\mathbf{r}}_i$ are the position operators for the electrons and protons, respectively.

To calculate $\hat{U}_{\kappa,\kappa'}$, we must integrate the spatial overlap of the photon modes $\kappa$ and $\kappa'$ with the charge density $\rho(\mathbf{r})$. This however may provide us with microscopic details of dynamics but would not be able to incorporate collective effects which are essential to observe novelty in plasma. So we take different approach to derive mode interaction function. We consider mode coupling constant and incorporate changes by factoring in interactions and coupling. Thus, the exact expression for $\hat{U}_{\kappa,\kappa'}$ is,

$$\hat{U}_{\kappa,\kappa'} = \frac{4\pi e^2}{\left(|\mathbf{q}|^2 + \frac{1}{\lambda_D^2}\right)} \left(1 - \frac{a^2|\mathbf{q}|^2}{1+a^2|\mathbf{q}|^2} \frac{2}{\lambda_D}\right). \tag{52}$$

This expression shows that the mode interaction function depends on the positions of the electrons and protons. $|\mathbf{q}|$ represents momentum transfer between photon modes. First term represents Debye shielded coulomb interaction. This complete structure of mode interaction function indicates towards intermediate range of interaction and collective effects which is not available in theories developed by Maxwell's electrodynamics. Long range interaction is suppressed by Debye shielding, very short ranges are suppressed by Podolsky factor. This results in interaction on surface of imaginary concentric hollow spheres from the charge being considered.

*2.2.4.3. Hamiltonian for terms quadratically dependent on Bosonic operators.* We now consider the total Hamiltonian for the system involving free electron and free proton/ion Hamiltonian both with minimal coupling, interaction Hamiltonians (electron-electron, electron-proton, proton-proton), Hamiltonian for the electromagnetic field in the Podolsky framework. This gives us the general form of total Hamiltonian as,

$$\hat{H}_{total} = \hat{H}_{free-electron} + \hat{H}_{free-proton} + \hat{H}_{interaction} + \hat{H}_{EM}. \tag{53}$$

For a system with photons and charged particles, the vector potential $\hat{\mathbf{A}}_{total}$ has been expressed in terms of photon modes isolated in equations as,

$$\hat{\mathbf{A}}_{total}(\mathbf{r},t) = \int \frac{d^3k}{(2\pi)^3} \sum_\lambda \mathbf{e}_{\mathbf{k},\lambda} \left[ \frac{\hat{a}_{\mathbf{k},\lambda}}{\sqrt{1+a^2\mathbf{k}^2}} e^{i(\mathbf{k}\cdot\mathbf{r}-\omega_{\mathbf{k},total}t)} + \frac{\hat{a}^\dagger_{\mathbf{k},\lambda}}{\sqrt{1+a^2\mathbf{k}^2}} e^{-i(\mathbf{k}\cdot\mathbf{r}-\omega_{\mathbf{k},total}t)} \right]. \tag{54}$$

In a plasma, photons can interact with each other through the charged particles, leading to a coupling between different photon modes $\kappa$ and $\kappa'$. This can be expressed through the interaction Hamiltonian. General form of interaction Hamiltonian can be written as,

$$\hat{H}_{int} = \sum_{\kappa,\kappa'} \Delta_{\kappa,\kappa'} \left( \hat{U}_{\kappa,\kappa'} \hat{a}_\kappa \hat{a}_{\kappa'} + \hat{U}^\dagger_{\kappa,\kappa'} \hat{a}^\dagger_\kappa \hat{a}^\dagger_{\kappa'} + h.c \right).$$

We make substitutions for $\Delta_{\kappa,\kappa'}$ and $\hat{U}_{\kappa,\kappa'}$. This gives us,

$$\hat{H}_{int} = \sum_{\kappa,\kappa'} \left( \frac{4\pi e^2}{\left(|\mathbf{q}|^2 + \frac{1}{\lambda_D^2}\right)} \right)^2 \left( 1 - \frac{1}{a^2} \frac{2\frac{1}{\lambda_D} - |\mathbf{q}|^2}{|\mathbf{q}|^2 + \frac{1}{\lambda_D^2}} \right) \frac{1}{\sqrt{(1+a^2|\mathbf{k}|^2)(1+a^2|\mathbf{k'}|^2)}}$$

$$\times \left( \left( 1 - \frac{a^2 |\mathbf{q}|^2}{1+a^2 |\mathbf{q}|^2} \frac{2}{\lambda_D} \right) \right) \left( \hat{a}_\kappa \hat{a}_{\kappa'} + \hat{a}^\dagger_\kappa \hat{a}^\dagger_{\kappa'} \right) + h.c. \quad (55)$$

The may be interpreted to indicate towards phenomena of photon condensation, and also indicates anisotropy in photon mode interactions due to wave vector difference term for different photon modes.

The final Hamiltonian,

$$\hat{H}_{plasma} = \sum_\kappa \omega_\kappa \hat{a}^\dagger_\kappa \hat{a}_\kappa + \sum_{\kappa,\kappa'} \Delta_{\kappa,\kappa'} \left( \hat{U}_{\kappa,\kappa'} \hat{a}_\kappa \hat{a}_{\kappa'} + \hat{U}^\dagger_{\kappa,\kappa'} \hat{a}^\dagger_\kappa \hat{a}^\dagger_{\kappa'} + h.c \right), \quad (56)$$

where, the first term $\sum_\kappa \omega_\kappa \hat{a}^\dagger_\kappa \hat{a}_\kappa$ represents the energy of the quantized photon field in the plasma, where each mode $\kappa$ has a modified frequency $\omega_\kappa$.

The term $\hat{a}_\kappa^\dagger \hat{a}_\kappa$ represents the number of photons in mode $\kappa$, contributing energy $\hbar\omega_\kappa$ per photon. In the plasma, the photon modes are influenced by the presence of charged particles and the Debye screening.

The mode-mode coupling terms $\hat{a}_\kappa \hat{a}_{\kappa'}$ and $\hat{a}_\kappa^\dagger \hat{a}_{\kappa'}^\dagger$ describe processes where photon pairs are created or annihilated. In the plasma, these processes are influenced by the collective plasma oscillations and the screening effects due to free electrons and ions. The coupling between photon modes, mediated by the plasma, introduces non-local interactions that lead to plasma waves and photon-plasma interactions. These are important for understanding wave-particle interactions in the plasma, such as Landau damping and plasma heating.

### 2.2.5. Bogoliubov transformation

We now utilize Bogoliubov transformation for Hamiltonian diagonalization [49-53]. Purpose of this is to simplify the dynamics by decoupling modes. The form of Bogoliubov transformation used here is,

$$\hat{a}_\kappa = \cosh(\theta_\kappa) \hat{b}_\kappa - \sinh(\theta_\kappa) \hat{b}_\kappa^\dagger, \tag{57}$$

and

$$\hat{a}_\kappa^\dagger = \cosh(\theta_\kappa) \hat{b}_\kappa^\dagger - \sinh(\theta_\kappa) \hat{b}_\kappa, \tag{58}$$

where, $\theta_\kappa$ is a parameter that depends on the photon mode, can be applied to simplify the Hamiltonian. This transformation introduces new bosonic operators $\hat{b}_\kappa$ and $\hat{b}_\kappa^\dagger$, which correspond to quasi-particles that diagonalize the Hamiltonian.

This transformation cause mixing of the creation and annihilation operators which in turn modifies the vacuum state of the photon field. The new operators $\hat{b}_\kappa$ represent quasi-particle excitations that account for photon pair creation or annihilation processes. $\theta_\kappa$ is chosen in a way to a describe the system where photons form a condensate (a Bose-Einstein condensate of photons). This happens where photon interactions dominate resulting in macroscopic occupation of certain photon modes. The Bogoliubov transformation would change the nature of the photon-photon interactions in the plasma. It softens the photon pair creation/annihilation terms, making them more tractable for further analysis. It also reveals new stable modes or instabilities in the photon field.

Application of the Bogoliubov transformation causes the Hamiltonian now to describe the energy levels and interactions of the new quasi-particles $\hat{b}_\kappa$ and $\hat{b}_\kappa^\dagger$ instead of old creation and annihilation operators. These quasi-particles correspond to the new vacuum state and excitations of the photon field in the plasma, capturing both the photon-photon interactions and the plasma-mediated effects. The general form of the Hamiltonian we are working with is of equation (56). To diagonalize this Hamiltonian, we apply the Bogoliubov transformation to the bosonic operators $\hat{a}_\kappa$ and $\hat{a}_\kappa^\dagger$, and $\theta_\kappa$ is the Bogoliubov angle to be determined.

We substitute the Bogoliubov-transformed operators into the Hamiltonian from equation (2.56). For the first term $\omega_\kappa \hat{a}_\kappa^\dagger \hat{a}_\kappa$, we get

$$\omega_\kappa \hat{a}_\kappa^\dagger \hat{a}_\kappa = \omega_\kappa \left( \cosh^2(\theta_\kappa) \hat{b}_\kappa^\dagger \hat{b}_\kappa - \sinh^2(\theta_\kappa) \hat{b}_\kappa \hat{b}_\kappa^\dagger + \cosh(\theta_\kappa)\sinh(\theta_\kappa) \left( \hat{b}_\kappa^\dagger \hat{b}_\kappa^\dagger + \hat{b}_\kappa \hat{b}_\kappa \right) \right) \qquad (59)$$

For the second term, $\Delta_{\kappa,\kappa'} \hat{U}_{\kappa,\kappa'} \hat{a}_\kappa \hat{a}_{\kappa'}$, we get,

$$\Delta_{\kappa,\kappa'}\hat{U}_{\kappa,\kappa'}\hat{a}_{\kappa}\hat{a}_{\kappa'} = \Delta_{\kappa,\kappa'}\hat{U}_{\kappa,\kappa'}\left(\cosh(\theta_{\kappa})\hat{b}_{\kappa} - \sinh(\theta_{\kappa})\hat{b}_{\kappa}^{\dagger}\right)\left(\cosh(\theta_{\kappa'})\hat{b}_{\kappa'} - \sinh(\theta_{\kappa'})\hat{b}_{\kappa'}^{\dagger}\right). \tag{60}$$

To diagonalize the Hamiltonian, we aim to eliminate the off-diagonal terms (such as $\hat{b}_{\kappa}^{\dagger}\hat{b}_{\kappa}^{\dagger}$ and $\hat{b}_{\kappa}\hat{b}_{\kappa}$) by choosing an appropriate value for $\theta_{\kappa}$. The Bogoliubov transformation is designed to mix the creation and annihilation operators in such a way that the off-diagonal terms vanish.

For the Hamiltonian to be diagonal, the coefficients of the off-diagonal terms (such as $\hat{b}_{\kappa}\hat{b}_{\kappa'}$) must vanish. This leads to the following condition for diagonalization,

$$\tanh(2\theta_{\kappa}) = \frac{2\Delta_{\kappa,\kappa'}\hat{U}_{\kappa,\kappa'}}{\omega_{\kappa} + \omega_{\kappa'}} \tag{61}$$

where, $\tanh(2\theta_{\kappa})$ gives the mixing angle for the photon modes. From the relation $\tanh(2\theta_{\kappa}) = \frac{\sinh(2\theta_{\kappa})}{\cosh(2\theta_{\kappa})}$, we can solve for $\cosh(2\theta_{\kappa})$ and $\sinh(2\theta_{\kappa})$,

$$\cosh(2\theta_{\kappa}) = \frac{\omega_{\kappa} + \omega_{\kappa'}}{\sqrt{(\omega_{\kappa} + \omega_{\kappa'})^2 - 4\Delta_{\kappa,\kappa'}^2\hat{U}_{\kappa,\kappa'}^2}}, \tag{62}$$

and

$$\sinh(2\theta_{\kappa}) = \frac{2\Delta_{\kappa,\kappa'}\hat{U}_{\kappa,\kappa'}}{\sqrt{(\omega_{\kappa} + \omega_{\kappa'})^2 - 4\Delta_{\kappa,\kappa'}^2\hat{U}_{\kappa,\kappa'}^2}} \tag{63}$$

After the diagonalization, the eigenvalues of the Hamiltonian correspond to the new photon energies in terms of the quasi-particle operators $\hat{b}_{\kappa}^{\dagger}\hat{b}_{\kappa}$. The eigenvalues $\lambda_{\kappa}$ are given by,

$$\lambda_\kappa = \sqrt{(\omega_\kappa + \omega_{\kappa'})^2 - 4\Delta_{\kappa,\kappa'}^2 \hat{U}_{\kappa,\kappa'}^2} \tag{64}$$

These eigen values represent the quasi-particle energies after the Bogoliubov transformation, and they describe the new excitation spectrum of the photon field in the plasma system. $\cosh(\theta_\kappa)$ and $\sinh(\theta_\kappa)$ describe the mixing between the photon creation and annihilation operators due to the mode-mode coupling. The values of these functions determine how much of the original photon modes are mixed in the new quasi-particle states. The eigen value indicates that the interaction reduces energy of the coupled modes while still retaining symmetry This increases stability and predicts conditions that may lead to photon condensation.

## 2.3. Total Hamiltonian

We now combine all the previously calculated components, i.e, free electron and proton Hamiltonians with minimal coupling to the electromagnetic field, electron-electron, proton-proton, and electron-proton (Coulomb interactions) with Debye screening, quantized electromagnetic wave Hamiltonian (electric and magnetic field components) in the Podolsky framework, photon-photon interactions mediated by charged particles, including pair creation/annihilation, podolsky regularization to prevent ultraviolet divergences and Bogoliubov transformation for diagonalization of the photon-photon interactions. This leads to the description of quasi-particles. Hamiltonian for Free Particles (Electrons and Protons) with Minimal Coupling

$$H_{free} = \sum_{i=1}^{Ne} \frac{1}{2m_e}(\hat{\mathbf{p}}_i - e\hat{\mathbf{A}}_i)^2 + \sum_{j=1}^{Np} \frac{1}{2m_p}(\hat{\mathbf{p}}_i - e\hat{\mathbf{A}}_i)^2. \tag{65}$$

Coulomb Interactions with Debye screening and Podolsky correction is represented as,

$$\hat{H}_{coulomb} = \frac{1}{2}\sum_{i\neq k}^{N_e} \frac{e^2}{|\hat{\mathbf{r}}_i - \hat{\mathbf{r}}_k|}\left(1 - \frac{|\hat{\mathbf{r}}_i - \hat{\mathbf{r}}_k|^2}{a^2}\right) e^{\frac{-|\hat{\mathbf{r}}_i - \hat{\mathbf{r}}_k|}{\lambda_D}} + \sum_{i=1}^{N_e}\sum_{j=1}^{N_p} \frac{e^2}{|\hat{\mathbf{r}}_j - \hat{\mathbf{r}}_i|}\left(1 - \frac{|\hat{\mathbf{r}}_i - \hat{\mathbf{r}}_j|^2}{a^2}\right) e^{\frac{-|\hat{\mathbf{r}}_i - \hat{\mathbf{r}}_j|}{\lambda_D}}$$

$$+ \frac{1}{2}\sum_{j\neq l}^{N_p} \frac{e^2}{|\hat{\mathbf{r}}_j - \hat{\mathbf{r}}_l|}\left(1 - \frac{|\hat{\mathbf{r}}_j - \hat{\mathbf{r}}_l|^2}{a^2}\right) e^{\frac{-|\hat{\mathbf{r}}_j - \hat{\mathbf{r}}_l|}{\lambda_D}},$$

(66)

where, $\hat{\mathbf{r}}_i$ and $\hat{\mathbf{r}}_j$ are the position operators for electrons and protons, $\lambda_D$ is the Debye length, accounting for the screening of the Coulomb interaction, $a$ gives Podolsky parameter with dimensions of length. This term includes electron-electron interactions where $i, j$ represent electrons. $i$ in electron-proton interactions represent an electron and $j$ represents a proton. Proton-proton interactions utilize both the indices to represent protons.

Electromagnetic field Hamiltonian in Podolsky framework as in equation (37),

$$H_{EM} = \frac{d^3k}{(2\pi)^3} \sum_\lambda \hbar \omega_{k,total} \left(\hat{a}^\dagger_{k,\lambda} \hat{a}_{k,\lambda} + \frac{1}{2}\right). \tag{67}$$

Photon-Photon Interactions (Including Bogoliubov Diagonalization) as obtained from equation (56),

$$\hat{H}_{photon-photon} = \sum_{\kappa,\kappa'} \Delta_{\kappa,\kappa'} \left(\hat{U}_{\kappa,\kappa'} \hat{a}_\kappa \hat{a}_{\kappa'} + \hat{U}^\dagger_{\kappa,\kappa'} \hat{a}^\dagger_\kappa \hat{a}^\dagger_{\kappa'}\right) + \sum_\kappa \lambda_\kappa \hat{b}^\dagger_\kappa \hat{b}_\kappa \tag{68}$$

By combining all these contributions, we can write the total Hamiltonian for the plasma system as,

$$\hat{H}_{total} = \hat{H}_{free} + \hat{H}_{coulomb} + \hat{H}_{EM} + \hat{H}_{photon-photon}. \tag{69}$$

The complete Hamiltonian gives us a detailed description of the plasma system permeated by photons which is compatible with Podolsky electrodynamics and quantum field theory. The expanded total Hamiltonian reveals how cross-group terms from the kinetic energy, photon-particle interactions, Coulomb interactions, and electromagnetic wave components combine to drive key collective phenomena in the plasma system.

## 2.4. Coupling Constant

We now proceed to identify and obtain an expression the plasma system Hamiltonian for the coupling operators. Coupling operator $C_\kappa$ relates to how matter fields (such as electrons and protons) couple to bosonic photon modes in a quantized framework. So, we need to account for electron-photon and proton-photon interactions (via minimal coupling), interaction of particles with the quantized electromagnetic field.

We consider the minimal coupling between the charged particles (electrons and protons) and the electromagnetic field, which in our Hamiltonian was given by,

$$\hat{H}_{\min imal} = -\sum_{i=1}^{N_e} e\hat{\mathbf{p}}_i . \hat{\mathbf{A}}(\hat{\mathbf{r}}_i, t) + \sum_{j=1}^{N_p} e\hat{\mathbf{p}}_j . \hat{\mathbf{A}}(\hat{\mathbf{r}}_j, t) + \frac{e^2}{2} \sum_{i=1}^{N_e} \hat{\mathbf{A}}^2(\hat{\mathbf{r}}_i, t).$$

(70)

This represents the interaction between the charged particles and the quantized electromagnetic field. The vector potential $\hat{\mathbf{A}}(\hat{\mathbf{r}}, t)$ can be written as a sum over the photon modes $\kappa$, where each mode is represented by a bosonic creation and annihilation operator in coherent state formalism,

$$\hat{\mathbf{A}}(\hat{\mathbf{r}}, t) = \sum_\kappa \mathbf{e}_\kappa \left( \frac{\hat{a}_\kappa e^{i\mathbf{k}.\hat{\mathbf{r}}}}{\sqrt{2\varepsilon_0 V \omega_\kappa}} + \frac{\hat{a}_\kappa^\dagger e^{-i\mathbf{k}.\hat{\mathbf{r}}}}{\sqrt{2\varepsilon_0 V \omega_\kappa}} \right).$$

(71)

Here $\hat{a}_\kappa$ and $\hat{a}_\kappa^\dagger$ are the annihilation and creation operators for the photon mode $\kappa$. $\mathbf{e}_\kappa$ is the polarization vector of the mode $\kappa$, $\omega_\kappa$ is the frequency of the photon mode $\kappa$ and $V$ is the normalization volume of the field. Substituting this into the minimal coupling term, we get the coupling between the particles and the photon field. For each particle (electron or proton), the coupling operator will involve the particle's momentum and the photon field.

We now identify the coupling operator $C_\kappa$ for our system. For an electron, the coupling term becomes,

$$C_{\kappa,e} = \sum_{i=1}^{N_e} \left[ \frac{1}{\sqrt{m_e \omega_\kappa}} \hat{p}_i \mathbf{e}_\kappa . \hat{u}_\kappa(\hat{\mathbf{r}}_i) + \frac{e}{\sqrt{2\varepsilon_0 V \omega_\kappa}} \hat{A}_\kappa \right], \tag{72}$$

where, $\hat{p}_i$ is the momentum operator for the $i^{th}$ electron. $\hat{u}_\kappa(\hat{\mathbf{r}}_i)$ is the mode function for the photonic mode $\kappa$ evaluated at the position of the electron. $\hat{A}_\kappa$ is the vector potential in terms of the creation and annihilation operators of the photon field. For a proton, the coupling operator $C_{\kappa,p}$ has the same form but with $m_p$ (the mass of the proton) and appropriate substitutions for proton-specific terms. Presence of $\omega_\kappa$ introduces Podolsky correction in the expression due to the modified dispersion relation calculated earlier.

The total coupling operator $C_\kappa$ for the plasma system (including both electrons and protons) can now be written as,

$$C_\kappa = \sum_{i=1}^{N_e} \left[ \frac{1}{\sqrt{m_e \omega_\kappa}} \hat{p}_i \mathbf{e}_\kappa . \hat{u}_\kappa(\hat{\mathbf{r}}_i) + \frac{e}{\sqrt{2\varepsilon_0 V \omega_\kappa}} \hat{A}_\kappa \right] + \sum_{j=1}^{N_p} \left[ \frac{1}{\sqrt{m_p \omega_\kappa}} \hat{p}_j \mathbf{e}_\kappa . \hat{u}_\kappa(\hat{\mathbf{r}}_j) - \frac{e}{\sqrt{2\varepsilon_0 V \omega_\kappa}} \hat{A}_\kappa \right]. \tag{73}$$

This expression includes both, the momentum coupling of each electron and proton to the photon modes, and the vector potential contribution (representing direct interaction with the photon field). $\hat{u}_\kappa(\hat{\mathbf{r}}_i)$ describe how the photon field varies across space. For plane waves, this would be $e^{i\mathbf{k}\cdot\hat{\mathbf{r}}_i}$ for mode $\kappa$. The coupling constant $c_\kappa$ for the interaction is determined by factors like $\frac{e}{\sqrt{2\varepsilon_0 V \omega_\kappa}}$, which controls the strength of the coupling between the particles and the photon modes. We however do not utilize coupling constant and have constructed coupling operator in manner to be sufficient for description of our system. Final expression for the plasma system coupling operator,

$$C_\kappa = \sum_{i=1}^{N_e} \frac{1}{\sqrt{m_e \omega_\kappa}} \hat{p}_i \mathbf{e}_\kappa \cdot \hat{u}_\kappa(\hat{\mathbf{r}}_i) + \sum_{j=1}^{N_p} \frac{1}{\sqrt{m_p \omega_\kappa}} \hat{p}_j \mathbf{e}_\kappa \cdot \hat{u}_\kappa(\hat{\mathbf{r}}_j) + \frac{\hat{A}_\kappa}{\sqrt{2\varepsilon_0 V \omega_\kappa}} (N_e e + N_p q_p). \tag{74}$$

This coupling operator encodes the interaction between the electron and proton momenta and the photon modes $\kappa$, as well as their interaction with the vector potential in the quantized electromagnetic field.

We now express complete Hamiltonian for the plasma system in terms of the coupling operator $C_\kappa$. The total Hamiltonian for the system can be expressed as the sum of the following components,

$$\hat{H}_{total} = \hat{H}_{free} + \hat{H}_{coulomb} + \hat{H}_{EM} + \hat{H}_{minimal} + \hat{H}_{photon-photon}.$$

We will now expand each term while incorporating the coupling operator $C_\kappa$.

The interaction Hamiltonian in the minimal coupling (equation 70) scheme is given by,

$$\hat{H}_{\text{minimal}} = \sum_{\kappa} C_{\kappa}\hat{a}_{\kappa} + C_{\kappa}^{\dagger}a_{\kappa}^{\dagger}, \tag{75}$$

where, $C_{\kappa}$ is the coupling operator for mode $\kappa$, describing the interaction between the particles and the quantized photon field. $\hat{p}_i$ and $\hat{p}_j$ are the momentum operators of the electrons and protons. $u_{\kappa}(\hat{\mathbf{r}}_i)$ and $u_{\kappa}(\hat{\mathbf{r}}_j)$ are the mode functions for the photonic field. This term describes how the charged particles interact with the photon field.

The complete Hamiltonian of the plasma system, incorporating all terms, can now be written as,

$$\hat{H}_{total} = \left(\sum_j \left(\frac{\hat{p}_{e_j}^2}{2m_e}\right) + \sum_j \left(\frac{\hat{p}_{p_j}^2}{2m_p}\right)\right) + \left(\sum_{\kappa} C_{\kappa}\hat{a}_{\kappa} + C_{\kappa}^{\dagger}a_{\kappa}^{\dagger}\right)$$

$$+ \left(\sum_{\kappa,\kappa'} \Delta_{\kappa,\kappa'}\left(\hat{U}_{\kappa,\kappa'}\hat{a}_{\kappa}\hat{a}_{\kappa'} + \hat{U}_{\kappa,\kappa'}^{\dagger}\hat{a}_{\kappa}^{\dagger}\hat{a}_{\kappa'}^{\dagger} + h.c\right)\right) + \sum_{\kappa} \omega_{\kappa}\hat{a}_{\kappa}^{\dagger}\hat{a}_{\kappa}. \tag{76}$$

This Hamiltonian provides a complete description of the plasma system in the quantized regime, with all relevant terms and interactions. Coulomb interactions and photon mode-mode interactions are contained in term with mode coupling constant and mode interaction function. If coupling strength constant is defined separately for a plasma system it would appear as pre-factor to minimal coupling term that is alongside coupling operator.

## 3. THE EFFECTIVE PLASMA MODEL

We aim to take Hamiltonian calculated in equation (76) and trace out photonic degrees of freedom in order to define complete system in terms of plasma degrees of freedom. We utilize coherent path integral approach [54] to define effective matter action which in our case is effective plasma action. The structure of Hamiltonian found in equation (76) indicates towards effective plasma-plasma coupling. This will appear in effective action. This will lead us to effective Hamiltonian in thermodynamic limit of $N \to \infty$.

### 3.1. Effective action: Euclidean path integral formulation

We consider Hamiltonian structure taken in equation (76) and then express the partition function as a path integral over the fields (both matter and photonic degrees of freedom). The total Hamiltonian for the plasma system in structure is,

$$\hat{H}_{total} = \hat{H}_{free} + \hat{H}_{coulomb} + \hat{H}_{minimal} + \hat{H}_{EM} + \hat{H}_{photon-photon}.$$

Our plasma system is a canonical ensemble for which the partition function $Z$ in quantum statistical mechanics is expressed as [55,56],

$$Z = Tr\left(e^{-\beta \hat{H}_{total}}\right), \tag{77}$$

where, $\beta = \dfrac{1}{k_B T}$ is the inverse temperature and $\hat{H}_{total}$ is the total Hamiltonian of the system. We can represent photonic field using coherent states $\hat{a}_\kappa |z_\kappa\rangle = z_\kappa |z_\kappa\rangle$. Using this form we can state the partition function for plasma system as,

$$Z = Tr_M\left(\int \prod_\kappa \frac{d^2 z_\kappa}{\pi} \langle z | e^{-\beta \hat{H}_{total}} | z \rangle \right). \tag{78}$$

This expression represents total partition function for plasma system that is permeated with photons. But for a specific expression we calculate individual expressions for partition function and then find their product to give total partition function of the system. This expression can be stated as,

$$Z = Z_{electron}.Z_{proton}.Z_{EM}.Z_{interaction}. \tag{79}$$

Considering the overlap and completeness condition for coherent state and expressing hamiltonian in terms of ladder operators we can write complete Hamiltonian as,

$$Z = \left(V\frac{2\pi m_e}{\beta}\right)^{\frac{3N_e}{2}} \left(V\frac{2\pi m_p}{\beta}\right)^{\frac{3N_p}{2}} \prod_\kappa \frac{1}{1-e^{-\beta\omega_k}} e^{-\beta\sum_{i\neq j}\frac{e^2}{4\pi\varepsilon_0|\mathbf{r}_i-\mathbf{r}_j|}\left(1-\frac{|\mathbf{r}_i-\mathbf{r}_j|^2}{a^2}\right)e^{-|\mathbf{r}_i-\mathbf{r}_j|/\lambda_D}}, \tag{80}$$

here, $V$ is the volume of the system, $N_e$ is number of electrons, $N_p$ is the number of proton, $M_e$ is mass of the electron, $M_p$ is the mass of proton and frequency is modified by Podolsky dispersion relation. For us to assign action to the total Hamiltonian of the system it is necessary that a coherent path integral partition function could be constructed. We transform coupling operator to field form. This coupling operator field and its conjugate can be written in the form,

$$C_\kappa = \frac{1}{\sqrt{\omega_\kappa}}\left[\sum_{i=1}^{N_e}\frac{\mathbf{p}_i.\mathbf{e}_\kappa}{\sqrt{m_e}}u_\kappa(\mathbf{r}_i) + \sum_{j=1}^{N_p}\frac{\mathbf{p}_j.\mathbf{e}_\kappa}{\sqrt{m_p}}u_\kappa(\mathbf{r}_j) + \frac{1}{\sqrt{2\varepsilon_0 V}}A_\kappa(N_e e + N_p q_p)\right], \tag{81}$$

$$\bar{C}_\kappa = \frac{1}{\sqrt{\omega_\kappa}}\left[\sum_{i=1}^{N_e}\frac{\mathbf{p}_i.\mathbf{e}_\kappa}{\sqrt{m_e}}u_\kappa^*(\mathbf{r}_i) + \sum_{j=1}^{N_p}\frac{\mathbf{p}_j.\mathbf{e}_\kappa}{\sqrt{m_p}}u_\kappa^*(\mathbf{r}_j) + \frac{1}{\sqrt{2\varepsilon_0 V}}A_\kappa^*(N_e e + N_p q_p)\right]. \tag{82}$$

We can also see that $z$ is equivalent of Fourier coefficient of vector potential written in equation (9). This eases the calculation and also makes inclusion of Podolsky correction in field form more coherent which may later be transformed back to quantum state as per the need. Total action can be written as sum of plasma action and action for multimode photonic field of the form,

$$S_\kappa = \int_0^\beta \left( \bar{z}_\kappa \partial_u z_\kappa + H_\kappa(z_\kappa, \bar{z}_\kappa) \right) du \qquad (83)$$

The above equation can be further be expanded to the form,

$$S_\kappa = \int_0^\beta \left( \bar{z}_\kappa \partial_u z_\kappa + \tilde{\omega}_\mathbf{k} \bar{z}_\kappa z_\kappa - c_\kappa \left( C_\kappa z_\kappa + \bar{C}_\kappa \bar{z}_\kappa \right) \right) du, \qquad (84)$$

Further substitution of Podolsky modified frequency and mode functions as coefficients of Fourier transform of vector potential gives,

$$S_\kappa = \int_0^\beta \left( \bar{A}_{\mathbf{k},\lambda} \partial_u A_{\mathbf{k},\lambda} + \frac{|\mathbf{k}|}{\sqrt{1+a^2 k^2}} \bar{A}_{\mathbf{k},\lambda} A_{\mathbf{k},\lambda} - c_\kappa \left( C_\kappa A_{\mathbf{k},\lambda} + \bar{C}_\kappa \bar{A}_{\mathbf{k},\lambda} \right) \right) du. \qquad (85)$$

First term in the above expression describes kinetic term for the mode $\kappa$ describing coherent state variables in this case coefficients of Fourier transform of vector potential over imaginary time $u$. Coupling constant can be omitted because there is no external coupling apart from field's inherent coupling which has already been accounted by coupling operator field in third term.

We can see that effective matter action can be defined by tracing over the light degrees of freedom,

$$Z = Tr_{plasma} \left( e^{-S_{plasma}} \prod_\kappa \int D(z_\kappa, \bar{z}_\kappa) e^{-S_\kappa} \right) = Z_0 Tr_{plasma} \left( S^{-eff} \right), \qquad (86)$$

Substituting the value of multimode photonic field,

$$Z = Tr_{plasma}\left(e^{-S_{plasma}} \prod_{\kappa}\int D(z_\kappa,\bar{z}_\kappa) e^{-\int_0^\beta \left(\bar{A}_{\mathbf{k},\lambda}\partial_u A_{\mathbf{k},\lambda} + \frac{|\mathbf{k}|}{\sqrt{1+a^2k^2}}\bar{A}_{\mathbf{k},\lambda}A_{\mathbf{k},\lambda} - c_\kappa(C_\kappa A_{\mathbf{k},\lambda}+\bar{C}_\kappa\bar{A}_{\mathbf{k},\lambda})\right)du}\right).$$ (87)

We need to calculate,

$$\prod_{\kappa}\int D(z_\kappa,\bar{z}_\kappa) e^{-\int_0^\beta \left(\bar{A}_{\mathbf{k},\lambda}\partial_u A_{\mathbf{k},\lambda} + \frac{|\mathbf{k}|}{\sqrt{1+a^2k^2}}\bar{A}_{\mathbf{k},\lambda}A_{\mathbf{k},\lambda} - c_\kappa(C_\kappa A_{\mathbf{k},\lambda}+\bar{C}_\kappa\bar{A}_{\mathbf{k},\lambda})\right)du}.$$ (88)

Quantized fields in thermal equilibrium, naturally give rise to periodic thermal trajectories. These can be mathematically stated as, $z(0)=z(\beta), \bar{z}(0)=\bar{z}(\beta)$. Fields expanded in Fourier series can be stated as,

$$z_\kappa(u) = \sum_{n=-\infty}^{\infty} z_{\kappa,n} e^{i\omega_n u},$$

and

$$\bar{z}_\kappa(u) = \sum_{n=-\infty}^{\infty} \bar{z}_{\kappa,n} e^{-i\omega_n u}.$$

Since we see that periodicity of our system is in boundary interval $[0,\beta]$, expansion should account for any inverse temperature dependence. This is done by term $\frac{1}{\beta}$, Without the inverse term is not restricted to any interval and maybe mathematically correct but thermodynamically lacks precision. Inverse term restricts periodicity to matsubara interval $[0,\beta]$ and directly reflects temperature dependence. This transforms the equation to,

$$z_\kappa(u) = \frac{1}{\sqrt{\beta}}\sum_n z_{\kappa,n} e^{i\omega_n u},$$ (89)

$$\bar{z}_\kappa(u) = \frac{1}{\sqrt{\beta}} \sum_n \bar{z}_{\kappa,n} e^{-i\omega_n u}, \tag{90}$$

where, $\omega_n = \dfrac{2\pi n}{\beta}$ are Matsubara frequencies [57-59].

$$S_\kappa = \sum_n \left( (i\omega_n + \tilde{\omega}_\kappa) \bar{z}_{\kappa,n}(\omega_n) z_{\kappa,n}(\omega_n) - \beta c_\kappa \left( C_\kappa z_{\kappa,n} + \bar{C}_\kappa \bar{z}_{\kappa,n} \right) \right), \tag{91}$$

where,

$$C_\kappa(\omega_n) = \frac{1}{\sqrt{\beta}} \int_0^\beta du \frac{1}{\sqrt{\omega_\kappa}} \left[ \sum_{i=1}^{N_e} \frac{\mathbf{p}_i \cdot \mathbf{e}_\kappa}{\sqrt{m_e}} u_\kappa(\mathbf{r}_i) + \sum_{j=1}^{N_p} \frac{\mathbf{p}_j \cdot \mathbf{e}_\kappa}{\sqrt{m_p}} u_\kappa(\mathbf{r}_j) + \frac{1}{\sqrt{2\varepsilon_0 V}} A_\kappa \left( N_e e + N_p q_p \right) \right](u) e^{i\omega_n u}, \tag{92}$$

$$\bar{C}_\kappa(\omega_n) = \frac{1}{\sqrt{\beta}} \int_0^\beta du \frac{1}{\sqrt{\omega_\kappa}} \left[ \sum_{i=1}^{N_e} \frac{\mathbf{p}_i \cdot \mathbf{e}_\kappa}{\sqrt{m_e}} u_\kappa^*(\mathbf{r}_i) + \sum_{j=1}^{N_p} \frac{\mathbf{p}_j \cdot \mathbf{e}_\kappa}{\sqrt{m_p}} u_\kappa^*(\mathbf{r}_j) + \frac{1}{\sqrt{2\varepsilon_0 V}} A_\kappa^* \left( N_e e + N_p q_p \right) \right](u) e^{-i\omega_n u}, \tag{93}$$

In Fourier space action becomes separable for each mode. Jacobian of Fourier transformation is unity, this enables us to substitute functional differential over change of temperature with system's evolution in equation (88) functional differential of frequency trajectories. This also indicates that volume element in path integral remains unchanged. The integral over each mode is a two dimensional Gaussian integral,

$$I_n = \int dz_{\kappa,n} d\bar{z}_{\kappa,n} e^{-\beta(i\omega_n + \tilde{\omega}_\kappa) \bar{z}_{\kappa,n} z_{\kappa,n}}. \tag{94}$$

Calculating Gaussian integral in complex plane we get,

$$I_n = \frac{\pi}{\beta(i\omega_n + \tilde{\omega}_\kappa)}. \tag{95}$$

This leads to,

$$\Pi_\kappa \int D(z_\kappa, \bar{z}_\kappa) e^{-S_\kappa} = \Pi_\kappa Z_\kappa \exp\left[\frac{c_k^2}{\beta \tilde{\omega}_\kappa} \int \int_0^\beta du du' C_\kappa(u) \bar{C}_\kappa(u) \frac{\omega_\kappa}{\beta} \sum_n \frac{e^{i\omega_n(u-u')}}{i\omega_n + \tilde{\omega}_\kappa + \frac{1}{\lambda_D^2}}\right], \quad (96)$$

With kernel of value,

$$K_\kappa(u-u') = \frac{\omega_\kappa}{\beta} \sum_n \frac{e^{i\omega_n(u-u')}}{i\omega_n + \tilde{\omega}_\kappa + \frac{1}{\lambda_D^2}}. \quad (97)$$

Kernel describes correlation between field trajectories at different times. This is in a sense analogous to propagator. Matsubara frequency indicates discrete thermal excitation of the plasma system. Thus now utilizing the above expression and equation (86) we can write the expression for effective action,

$$S_{eff} = S_{plasma} - \sum_\kappa \frac{c_k^2 \omega_\kappa}{\tilde{\omega}_\kappa \sum_n \left(i\omega_n + \tilde{\omega}_\kappa + \frac{1}{\lambda_D^2}\right)} \int_0^\beta d\tau \int_0^\beta du' \left[C_\kappa \bar{C}_\kappa\right] e^{i\omega_n \tau}. \quad (98)$$

The expression includes kernel which encodes non-locality through correlations at different imaginary times and interaction term shows how plasma particles interact across different points in imaginary time. This shows that it cannot be directly be mapped to Hamiltonian for plasma subsystem. It also shows that Debye screening and Podolsky interaction can lead to non-local effects in imaginary time. For large number of particles the system becomes dominated by collective, time independent modes. This recovers time local interactions, where non-local kernels reduce to delta function eliminating imaginary time correlations. Non-locality arises because of the memory effects introduced by the environmental degrees of freedom.

## 3.2. Effective hamiltonian

We intend to derive effective Hamiltonian of the system in this section. This requires the scaling of plasma system. Number of plasma particles tending towards infinity is considered and as we know photonic operators scale as $\sqrt{N}$ and can be expressed as $b/\sqrt{N}$ and $b^{\dagger}/\sqrt{N}$ [60]. This gives us,

$$\int \prod_{\kappa} \frac{d^2 z_{\kappa}}{\pi} \langle z | e^{-\beta \hat{H}} | z \rangle = \int \prod_{\kappa} \frac{d^2 z_{\kappa}}{\pi} e^{-\beta \langle z | \hat{H} | z \rangle}, \qquad (99)$$

Where,

$$\langle z | \hat{H} | z \rangle = \hat{H}_{plasma} + \sum_{\kappa} \tilde{\omega} |z_{\kappa}|^2$$

$$+ \sum_{\kappa} c_{\kappa} \left( \frac{1}{\sqrt{\beta}} \int_0^{\beta} du \frac{1}{\sqrt{\omega_{\kappa}}} \left[ \sum_{i=1}^{N_e} \frac{\mathbf{p}_i \cdot \mathbf{e}_{\kappa}}{\sqrt{m_e}} u_{\kappa}^*(\mathbf{r}_i) + \sum_{j=1}^{N_p} \frac{\mathbf{p}_j \cdot \mathbf{e}_{\kappa}}{\sqrt{m_p}} u_{\kappa}^*(\mathbf{r}_j) + \frac{1}{\sqrt{2\varepsilon_0 V}} A_{\kappa}^*(N_e e + N_p q_p) \right] (u) e^{-i\omega_n u} z_{\kappa} + h.c. \right). \quad (100)$$

The kernel for the plasma system described in equation (97) includes Matsubara summations. Setting $\omega_n = 0$ in thermodynamic limit simplifies to,

$$K_{\kappa}(u-u') \to \delta(u-u') \qquad (101)$$

This eliminates temporal correlations. For large $N$ photonic fluctuations also vanish. Let us restate the kernel,

$$K_{\kappa}(u-u') = \frac{\omega_{\kappa}}{\beta} \sum_n \frac{e^{i\omega_n(u-u')}}{i\omega_n + \tilde{\omega}_{\kappa} + \frac{1}{\lambda_D^2}}$$

We define,

$$\Omega_\kappa = \tilde{\omega}_\kappa + \frac{1}{\lambda_D^2}. \tag{102}$$

Thus re-writing the kernel in the specific form utilizing above compactification and changing the form we get,

$$K_\kappa(u-u') = \frac{\omega_\kappa}{\beta} \sum_n \left( \frac{\Omega_\kappa e^{i\omega_n(u-u')}}{\omega_n^2 + \Omega_\kappa^2} - \frac{i\omega_n \omega_\kappa e^{i\omega_n(u-u')}}{\omega_n^2 + \Omega_\kappa^2} \right). \tag{103}$$

We can denote the above equation as,

$$K_\kappa(u-u') = K_\kappa^{(1)}(u-u') - iK_\kappa^{(2)}(u-u'), \tag{104}$$

where,

$$K_\kappa^{(1)}(u-u') = \frac{\omega_\kappa}{\beta} \frac{\Omega_\kappa e^{i\omega_n(u-u')}}{\omega_n^2 + \Omega_\kappa^2}, \tag{105}$$

and,

$$K_\kappa^{(2)}(u-u') = \frac{\omega_\kappa}{\beta} \sum_n \frac{\omega_n e^{i\omega_n(u-u')}}{\omega_n^2 + \Omega_\kappa^2}. \tag{106}$$

Splitting first term in delta functions we can write,

$$K_\kappa(u-u') = \omega_\kappa \Omega_\kappa \left[ \beta \sum_m \delta(\tau - m\beta) - \bar{\bar{K}}_\kappa(\tau) \right] - \frac{\partial_\tau}{\Omega_\kappa} \left[ \bar{K}_\kappa(\tau) \right], \tag{107}$$

and,

$$\bar{K}_\kappa(\tau) = \frac{\omega_\kappa}{\Omega_\kappa}\left[\beta\sum_m \delta(\tau - m\beta) - \bar{\bar{K}}_\kappa(\tau)\right], \tag{108}$$

Where,

$$\bar{\bar{K}}_\kappa(\tau) = \sum_n \frac{e^{i\omega_n \tau}}{\omega_n^2 + \Omega_\kappa^2}. \tag{109}$$

$\omega_\kappa$ in equation (107) represents unmodified standard frequency and $\Omega_\kappa$ represents Podolsky modified frequency and Debye shielding. First term arises from the Matsubara summation which reflects the periodic boundary conditions in imaginary time for the system in thermal equilibrium. The delta functions represent discrete points in imaginary time where correlations are maximized. These delta functions capture the thermal response of the plasma system, reflecting periodic correlations between particle and field states due to the system's temperature. The $\bar{\bar{K}}_\kappa(\tau)$ term represents finite-temperature corrections to the kernel, capturing the contributions of Matsubara frequencies with modified plasma dynamics. This term captures deviations from thermal delta-like correlations due to the modified dispersion relation and screening. It governs intermediate and long-range correlations in the plasma system. It encodes both collective and individual particle dynamics, blending thermal and quantum effects. Plasma oscillations are regularized by Podolsky modifications. The group velocity of electromagnetic waves is altered by $\Omega_\kappa$ affecting energy propagation in the plasma. $-\frac{\partial_\tau}{\Omega_\kappa}\left[\bar{K}_\kappa(\tau)\right]$ term originates from the imaginary-time evolution of the kernel. It reflects the coupling between temporal changes in particle trajectories and the plasma-modified photonic modes. It captures non-local temporal correlations, highlighting memory effects within the plasma system. It represents how the plasma system retains information about interactions over time. The term suggests delayed responses in plasma oscillations due to temporal correlations. It also introduces non-Markovian memory effects, where the plasma's behaviour depends on its past states. In equation (108) dominance of first term as $N \to \infty$ indicates time local interaction. This suggests that as the number of particles become large the kernel becomes time-local, enabling a Hamiltonian-based description. This reduces computational complexity in numerical simulations.

Second term $\bar{\bar{K}}_\kappa(\tau) = \sum_n \frac{e^{i\omega_n \tau}}{\omega_n^2 + \Omega_\kappa^2}$ becomes more prominent when environment is strongly interacting. This term arises from the finite-temperature contributions of non-zero Matsubara frequencies. Non-local effects reflect memory-like behaviour in the plasma's response to electromagnetic fields. These effects dominate in strongly interacting regimes or at finite temperatures, where the photon-plasma coupling leads to significant delays or persistent correlations. Non-local behaviour becomes very significant when number of plasma particles is finite. Non local behaviour indicates towards phenomena like photon condensation, continual non-local plasma correlation, etc. So for $N \to \infty$, effective action becomes local in time, giving us expression,

$$S_{eff} = S_{plasma} - \sum_\kappa \frac{\omega_\kappa}{\omega_\kappa + \frac{1}{\lambda_D^2}} \int_0^\beta dv \, C_\kappa(v) \bar{C}_\kappa(v). \tag{110}$$

Effective Hamiltonian resulting from this action is,

$$\hat{H}_{eff} = \hat{H}_{plasma} - \sum_\kappa \frac{\omega_\kappa}{\Omega_\kappa} \hat{C}_\kappa \hat{C}_\kappa^\dagger. \tag{111}$$

The effective Hamiltonian derived for the plasma system reveals that different behaviour for different densities is expected in plasma. For finite system temporal correlations and non-Markovian behaviour are dominant and as system becomes very large, temporal correlations become more local. This also reveals increased stability of plasma and collective behaviour.

## 4. PHOTON OBSERVABLES

We have derived effective Hamiltonian for the plasma system in previous section by eliminating photonic degrees of freedom. This however has been done in a manner that information for photonic field is not lost irrecoverably but can be reconstructed on the basis of information of plasma dynamics. Making few adjustments we can easily calculating photon displacement operator, photon current operator and photon number operator. We start with the Hamiltonian of full system,

$$\hat{H}_{total} = \hat{H}_{plasma} + \sum_{\kappa} \tilde{\omega}_{\kappa} \hat{b}^{\dagger}_{\kappa} \hat{b}_{\kappa} - \sum_{\kappa} \frac{\omega_{\kappa}}{\Omega_{\kappa}} \left( \hat{C}_{\kappa} \hat{b}_{\kappa} + h.c. \right), \tag{112}$$

where, $\hat{H}_{plasma}$ is isolated plasma particle Hamiltonian in isolation, and $\hat{H}_{total}$ is Hamiltonian of the complete system. Interaction term from equation (111) can be stated as,

$$-\sum_{\kappa} \frac{\omega_{\kappa}}{\Omega_{\kappa}} \left( \hat{C}_{\kappa} \hat{C}^{\dagger}_{\kappa} \right). \tag{113}$$

it describes the effective interaction mediated by the photonic field. From equation (111) and (112) we can see,

$$\left\langle \hat{b}_{\kappa} \right\rangle = -\frac{\omega_{\kappa}}{\Omega_{\kappa}} \left\langle \hat{C}_{\kappa} \right\rangle_{plasma}, \tag{114}$$

$$\left\langle \hat{b}^{\dagger}_{\kappa} \right\rangle = -\frac{\omega_{\kappa}}{\Omega_{\kappa}} \left\langle \hat{C}^{\dagger}_{\kappa} \right\rangle_{plasma}. \tag{115}$$

This in turn gives us photon displacement operator or otherwise known as photon field expectation value,

$$\langle b_\kappa + b_\kappa^\dagger \rangle = -\frac{\omega_\kappa}{\Omega_\kappa} \langle C_\kappa + C_\kappa^\dagger \rangle_{plasma}. \tag{116}$$

Substituting the values of coupling operators we get,

$$\langle b_\kappa + b_\kappa^\dagger \rangle = -\frac{\sqrt{\omega_\kappa}(N_e e + N_p q_p)}{\Omega_\kappa \sqrt{\varepsilon_0 V}} n_B(\beta, \Omega_\kappa). \tag{117}$$

Photon current operator comes out to be,

$$\langle b_\kappa - b_\kappa^\dagger \rangle = -\frac{\omega_\kappa}{\Omega_\kappa} \langle C_\kappa - C_\kappa^\dagger \rangle_{plasma}. \tag{118}$$

Since, $\langle C_\kappa - C_\kappa^\dagger \rangle_{plasma} = 0$ the value of photon current operator also turns to zero.

Photon number operator results into,

$$\langle b_\kappa^\dagger b_\kappa \rangle = n_B(\beta, \Omega_\kappa) + \frac{\omega_\kappa^2}{\Omega_\kappa^2} \langle C_\kappa C_\kappa^\dagger \rangle_{plasma}, \tag{119}$$

$$\langle b_\kappa^\dagger b_\kappa \rangle = n_B(\beta, \Omega_\kappa)\left(1 + \frac{\omega_\kappa^2}{\Omega_\kappa^2} \frac{(N_e e + N_p q_p)^2}{\omega_\kappa \varepsilon_0 V}\right). \tag{120}$$

Photon number operator can be translated to,

$$n_B(\beta, \Omega_\kappa) = \frac{1}{\left(1 + \frac{\omega_\kappa^2}{\Omega_\kappa^2} \frac{(N_e e + N_p q_p)^2}{\omega_\kappa \varepsilon_0 V}\right)}. \tag{121}$$

The coupling operator expectation reflects the induced polarization of the plasma due to the interaction with the photonic field. Photon displacement operator reflects the displacement of the photonic field due to the plasma's polarization response. Podolsky corrections and Debye screening modulate the coupling strength. The current operator describes the anti-symmetric response of the photonic field to plasma oscillations. Podolsky corrections and Debye screening ensure the response is regularized at short and long distances, respectively. The photon number operator predicts that Bose-Einstein terms dominate at high temperatures, while light-matter coupling terms become more relevant at low temperatures. Photon population deviates from original value because of plasma mediated photonic interactions and contributions. These contributions are inversely proportional to volume of the system. The coupling term reflects collective plasma oscillations, enhancing photon population in strongly interacting regimes. The plasma contribution increases with the net charge and photon frequency.

## 5. RESULTS

We have developed a model for plasma system permeated by photonic field utilizing generalized quantum electrodynamics (using Podolsky's theory of electrodynamics). We find two dispersion modes for electromagnetic wave and construct an effective dispersion relation which will account for resultant of both modes. Application of the theory reduces the contribution of high-momentum modes, leading to regularization of the field energy. Finite length scale of the characteristic length scale $a$ introduces a cutoff, preventing the ultraviolet divergences typical in classical electrodynamics. Structure of mode coupling constant predicts modified plasma oscillations due to modified screening. This also predicts enhanced stability especially in case of high energy and high density plasma. Dependence on $|\mathbf{q}|^2$ also suggests anisotropic behaviour in mode coupling. This directional dependence may lead to topological effects which have not previously been predicted. Podolsky parameter also suppresses small scale fluctuations. Dependence on wave vectors in denominator may facilitate mode locking, energy cascades and wave synchronization. This also points towards screened plasma clusters in certain conditions. Mode interaction function indicates towards intermediate range of interaction and collective effects which is not available in theories developed by Maxwell's electrodynamics. Long range interaction is suppressed by Debye shielding, very short ranges are suppressed by Podolsky factor. This results in interaction on surface of imaginary concentric hollow spheres from the charge being considered. The Eigen value of Bogoliubov transformations indicates that the interaction reduces energy of the coupled modes while still retaining symmetry. This increases stability and predicts conditions that may lead to photon condensation.

We also find non-Markovian behaviour of plasma system for finite number of plasma particles that is it retains memory of plasma states over a period of time. Enhanced stability, and plasma wave correlations over long spatial and temporal range has also been predicted. Delayed responses in plasma oscillations due to temporal correlations have also been predicted.

## Declaration of competing interest

The authors declare that they have no known competing financial interests or personal relationships that could have appeared to influence the work reported in this paper.

## Acknowledgement

The authors thank SERB- DST, Govt. Of India for financial support under MATRICS scheme (grant no. : MTR/2021/000471).